# "Because math": Epistemological stance or defusing social tension in QM?

Erin Ronayne Sohr[1], Benjamin W. Dreyfus[1], Ayush Gupta[1], and Andrew Elby[2]

[1]*University of Maryland, Department of Physics*
[2]*University of Maryland, Department of Teaching and Learning, Policy and Leadership*
*College Park, Maryland, 20742*

**Abstract.** In collaborative small-group work, physics students need to both manage social conflict and grapple with conceptual and epistemological differences. In this paper, we document several outlets that students use as tools for managing social conflict when addressing quantum mechanics tutorials in clinical focus groups. These resources include epistemic distancing, humor, playing on tutorial wording and looking ahead to subsequent questions. We present preliminary analysis of episodes where students work through a Particle in a Box tutorial. Each episode highlights a different manner of navigating social tension: through shared epistemic humor in one case, and reinterpretation of the question in the other.

## I. INTRODUCTION

Collaborative active learning using research-based materials can have many benefits. [1-3] Group learning allows students to share knowledge as they present and critique each other's ideas and reasoning strategies. Students working in these small groups have the potential to exhibit better problem solving solutions than when working individually. [2] Such settings could therefore help students who are learning to negotiate the boundary between quantum mechanics (QM) and classical mechanics. However, in addition to the difficulty of learning new concepts that are remote from intuitive reasoning [4-6], negotiating these concepts in collaborative group settings can create further challenges. [7, 8]

As a result, collaborative learning of QM (and other topics) can give rise to social conflict for a variety of reasons. The ideas introduced in the group and their connection to the end goal may be unclear. What is taken to be understood by the group fluctuates quickly; common ground is fluid and therefore demands constant attention by participants [3]. Conflicts may also arise from dominant personalities [9], unequal opportunities to participate [10], and failure to obey turn-taking norms or students' insistence on their own strategies. [3] Previous research has shown students using epistemic distancing [11] or slipping into less collaborative modes [3] as ways of mitigating tension that results from these interactional differences.

In this paper, we further this line of research in documenting new resources that students may use to navigate this type of tension. In particular, our data focus on "escape hatches"—ways to sidestep rather than resolve a substantive debate—that students find in playful talk, the wording of the tutorial, or appeals to math. Playful talk can include humor, hedging, or even teasing. [10]

Our main goal in this paper is to document some "escape hatches" by which social tension is relieved in small-group work and show how these discursive moves can act as conversational pivots, redirecting or terminating group discussions. In the first two episodes, we illustrate how a group's reliance on math without parsing the meaning of the mathematical expressions, while appearing at first glance to reflect an unproductive epistemological stance, actually arises as an escape hatch that might not reflect the individual students' epistemological views. In the third episode, we show how the wording of the tutorial afforded to the students an escape hatch from the social tension arising from their debate. In our discussion section, we recount other escape hatches that we have seen in our data. We then argue that these results can inform instructors' facilitation of small-group work below.

## II. METHODOLOGY

We are currently developing QM tutorials designed to help students more consciously select between quantum, classical, or hybrid reasoning when problem solving. The data presented here come from a focus group of students working through our Particle in a Box (PIAB) tutorial.

When viewing the video of the students working through the PIAB tutorial, we noticed students getting to tense points in their discussions and subsequently taking an escape hatch to avoid situations of social conflict. We inferred tension from a combination of tone of voice, changes in pacing of talk, body posture, and the substance of students' utterances. Through repeated viewings, we tried to identify and characterize all the escape hatches students used. These included the use of humor, epistemic or affective distancing, framing moves, and various appeals to math. [11]

After viewing the video, we selected several episodes that provide strong examples of students drawing on these resources to relieve tension. We are do not intend to make generalizations based on the small sample of data, but simply wish to document the existence of these resources and how their instantiation in these episodes is coupled with the conceptual content of the unfolding conversation.

The episodes discussed below come from a group of physics majors in the first semester of a QM course. The students were familiar with the PIAB and knew each other through other common course(s). The first episode comes from a clip where students discuss the question "Why isn't the ground state n=0? That is, why isn't it possible for the particle to have zero energy?" Here we see evidence of students deploying math in a way that defuses the social tension they are experiencing. The second episode comes from a section of the tutorial where the students are asked, "Can we define a 'speed' for the wave?" referring to a classical standing wave on a string as a model for an energy eigenstate of the PIAB. Here, the students engage in a long period of tense reasoning before defusing the tension by taking an escape hatch afforded by the wording of the question. We chose these episodes as demonstrations of two escape hatches because the talkative nature of the group made it possible to illustrate these escape hatches in detail. In unpacking the flow of events by which the escape hatches arose, we aim to begin exploring this phenomenon.

### III. RESULTS

#### A. Episode 1: Why can't a particle have zero energy?

In episode 1 (see http://hdl.handle.net/1903/16748 for transcript and PIAB tutorial), the students, whose pseudonyms are Al, Bob, Chad, Dan and Ed, begin their discussion with Chad reading the question to the group, "Why isn't it possible for the ground state to have zero energy?" Al immediately provides an answer when he suggests "Uncertainty principle." After some pause, where the other group members do not acknowledge his contribution, he continues with "I guess mathematically, I don't know why." We see this statement as opening the space for the group to try other approaches, particularly mathematical ones (which he implies are different from his "Uncertainty Principle" idea). The group takes up Al's suggestion by exploring various mathematical tools they have.

Tension begins to build between the group members as they take a detour into linear algebra. In response to Al's opening bid to explore mathematical resources, Dan starts with "the difference between that and the harmonic oscillator" in terms of how each energy scales with quantum number $n$. Al quickly corrects him with "right, but when you have like the state n=0 for a harmonic oscillator, you still like, in your equation for the energy levels, you still use n=1. You just call it n=0." Chad and Dan abruptly counter Al's claims, both talking at once and over Al as he concludes. Chad says "well, no. It's n=0 but it's n plus one half." Responding to Chad and Dan's quick criticism of his proposal, Al sounds defensive as he counters with "yeah, yeah, yeah. You're right but for like these, like the square well or whatever, there's an n multiplied by it so like, if you had n=0, the energy would be zero."

The group continues searching for a mathematical reason why the PIAB cannot have zero energy. "You can't have a particle with no energy. That's like saying I have a whole bushel of no apples," Chad says. Al chastises him with a demeaning "no" followed by an analogy to a classical ball in a well, during which Bob tries to hide a smirk behind his hand while exchanging looks with other members of the group. The tense discussion continues the detour into linear algebra following Bob's suggestion that "isn't there some sort of theorem in linear algebra that says that zero can't be an eigenvalue or is it an eigenvector?" Al immediately responds with authority that it's an eigenvector, but subsequently, yet confidently, changes his position to eigenvalue upon protests by Chad and Dan. Bob asks for clarification "oh, so it can't be an eigenvalue?" To which, Al delivers another chastising "no, no" with his disagreement. Bob tries to synthesize the group's position on the argument when Al interrupts him with "it's whichever one makes it trivial." The group comes to an agreement, deciding that zero can't be an eigenvalue,[1] and a joke by Al relieves the tension of the discussion and restores group cohesion. "So we can say linear algebra," Al says and the group relaxes and laughs. Dan agrees, saying "because math."

The "because math" joke reiterates the group's epistemic stance that mathematics is the preferred place to look for warrants for their arguments while also acknowledging, through humor, that their mathematical "resolution" is perhaps not fully satisfying. This is supported by Al's next comment "no, what I was arguing at the very beginning though, I thought at least qualitatively it boiled down to the uncertainty principle." Al's emphasis on "qualitative" suggests that he might be thinking of his reasoning based on the uncertainty principle as distinct from the mathematical reasoning they have been pursuing for the last few minutes. The group shows their support for the need for a conceptual response by allowing Al to complete this relatively long statement without interruption, after which they collectively explore what the uncertainty principle may offer. Hence, through the use of epistemological humor—humor *about* the types of knowledge they are drawing upon and constructing—the group transactionally constructs an epistemic stance that positions math as a primary, but not complete, intellectual resource. The verbal pivot "because math" serves two purposes: it solidifies this positioning of math as well as temporarily relieving social tension, allowing the group discussion to proceed in a new, potentially productive direction.

---

[1] We recognize that this is an incorrect conclusion, and that it's an eigenvector that can't be zero. However, if zero were not an allowed eigenvalue, then the ground state energy would not be allowed to be zero, from a mathematical standpoint.

## B. Episode 2: Measurements of Energy

To be clear, the group does not *always* display this epistemic stance toward math. In a different clip, we see evidence of the group supporting a slightly different epistemic stance towards math when discussing what values a measurement of energy could yield and whether a repeat measurement would yield that same value. Here, Dan and Ed immediately turn to math, searching for the equation for the allowed energy levels. Chad seems to see less value in reporting the exact equation, and jokingly tells Dan to "just say it's $E_n$." However, the full equation for $E_n$ seems important to Dan and Ed, who are now joined by Al and Bob in trying to find it. After some discussion on the possibility of evolving between states, Al summarizes the group's findings as "so we're saying you get the energy of the ground state, and you would always get that," to which the rest of the group agrees. To Al, this seems to be a satisfactory explanation; he writes his answer down.

However, this qualitative answer is insufficient for Ed and Dan. After Al's summary, Ed asks the group a second time for the energy equation, to which Chad again jokes that it's "$E_0$," while Dan replies, "I'm just saying it's equal to $E_1$." Chad concedes that it would be $E_1$ and not $E_0$, because it's a PIAB and not a harmonic oscillator. Bob, Chad, and Dan work together to remember the energy equation for the ground state of the PIAB. Once they agree upon the equation, Chad remarks, "seems good," which we interpret to indicate that Chad believes the group has come to a satisfactory answer.

In this segment, the group no longer treats math as a primary but not complete resource. Instead, students take more individualized epistemic stances. Specifically, in these moments, Al treats a qualitative answer (in terms of the "ground state") as sufficient. Chad seems to initially agree with Al, his joking manner hinting that the mathematical formula was unnecessary. However, Chad indulges Ed's insistence on finding the formula and seems satisfied at the mathematical result. Dan and Ed, by contrast, treat math as not only a primary but fully sufficient as an answer, as evidenced by their wanting to write the full equations for $E_n$. Interestingly, with his "because math" comment in the previous episode, Dan became a vocal supporter of the group's earlier epistemic stance that considered math primary but not complete.

Our point here is that the earlier epistemic stance cannot be attributed to the group as a stable, robust belief; in other episodes individual students (including Dan) take stances that differ both from each other and from that earlier stance. This indirectly supports our argument that the earlier stance was constructed by the group in the moment, partly as an escape hatch. The view that epistemic stances can take on context-dependent [12, 13] and transactional [14, 15] forms has seen growing support in the literature.

## C. Episode 3: Measurements of Speed

The second example of an escape hatch comes from the section of the tutorial where students are asked to make an analogy between the PIAB and a classical standing wave on a string. The question asks, "Can we define a 'speed' for the wave?" The group begins by discussing whether phase or group velocity would be appropriate, with qualitative descriptions of each. Bob reflects that "a standing wave is equivalent of two waves moving in opposite directions, superimposed on each other" to which Chad suggests that the speed of the standing wave could then be defined as that of either of those waves since they propagate at the same speed. Al disagrees, saying that he can imagine a scenario in which one component is travelling at a different speed than the other. Bob, Chad, and Dan then address Al, all talking at once. They each try to correct him and remind him that they're talking about a standing wave. Al shows evidence of feeling the tension of the group ganging up on him by defensively responding with "no, I get that" while he nervously taps on the table. Chad reiterates to Al that "it must be a standing wave." Al tries to reestablish his ground and starts with "No I get what you're saying, but..." His agitation is clear in his nervous drumming on the table and his rush to find the words to reestablish his position. Meanwhile Bob and Chad share in the social tension of this interaction, smirking at Al and exchanging glances.

The group continues to struggle over defining the speed of the wave. Al claims that he doesn't "think there is one," which leads Chad to check the tutorial, perhaps to see if this answer is acceptable—which it is! In a relieved tone, he remarks that the question reads "*can* we define," to which the group agrees that they cannot. Here, Chad has found an escape hatch in the wording of the tutorial, specifically in the use of the word "can" that allowed the group to relieve the tension they were experiencing from their discussion. However, relieving that social tension also meant a discontinuation of what some may see as a fruitful discourse on the physics of standing waves. (The lack of fruitfulness, in our view, stems not from students reaching a wrong or incomplete answer, but rather, from settling upon this conclusion before debating it more substantively.)

## IV. DISCUSSION

The "*can* we define" and "because math" escape hatches described above are just two examples. We have seen many others across multiple groups, including looking ahead to subsequent questions to look for unintentional scaffolding, playing on the use of quotation marks, not explaining or discussing an answer unless explicitly prompted by the tutorial, or superficially using consistency questions. Consistency questions embedded in the tutorial ask if responses to two previous questions are compatible, and students sometimes look ahead to use the question as an announcement that they should give consistent answers to the prior questions (but without thinking through *why* those answers are consistent). Finally, we saw a group, when faced with growing social tension, reframe the small-group activity as working side-by-side but not needing to agree upon answers—a permanent "escape."

Not all tension-relieving discourse moves are escape hatches. For instance, a tension-relieving outlet we and others [11] have observed, which often functions to facilitate rather than sidestep substantive discussion, is humor. At some points we see Chad read the questions in a silly or joking manner, perhaps as a bid to raise and answer the question "How seriously are we taking this?" We see this move as a preemptive escape hatch, a move that attempts to frame the activity as one where tension need not arise from discussion. We have also seen epistemic distancing [11], where students hedge their claims by portraying themselves as simply the messenger, not the author of their claim; and affective distancing, where students propose an idea that is not meant to be taken too seriously. We observe that these forms of distancing facilitate rather than derail discussions by allowing participants to engage in argumentation while at the same time alleviating the need to save face if the group reacts negatively to their idea.

## V. CONCLUSION AND IMPLICATIONS

In our data, we have students working in a conceptually difficult area of physics, the area where classical physics meets quantum physics. Negotiating reasoning in a group setting often leads to group conflict or tension through the difficulty of the subject matter, disagreements on how to approach a problem, interpersonal issues, or differences on what resources should be relied on.

In episodes not presented in this paper, we see students employ tension-avoidance and tension-reduction moves documented in previous literature, such as epistemic or affectual distancing [11], playful talk [10], or framing an activity in a less coordinated manner. [3] In this paper, we focused on a class of strategies which we called "escape hatches," in which students side-step rather than resolve conflicts.

Although it can be uncomfortable for students, conflict engenders cognitive development and shared understanding. [16-18] It is a responsibility of an instructor to ensure that these conflicts lead to meaningful experiences, which may mean intervening in these tense moments or not. In the "because math" episode, the joke serves as a verbal pivot in the conversation which allows the group to explore different resources. If instructors see this type of reasoning as productive, this episode may not have warranted an intervention. In the second episode, the "*can* we" distinction also serves as a conversational pivot; however it serves to escape the preceding debate rather than try to resolve it using different resources. In this episode, an instructor intervention may be more warranted than in the previous episode, where the escape hatch allowed the debate continue from a new angle, rather than squelching it completely. Our research highlights that these small-group settings are worthy of careful investigation by the instructor, in terms of the social dynamics of the group, the content of their discussion, and the interaction of the two. If left unchecked, tense moments arising from these interactions can become turning points in student reasoning, where students stray unproductively far from what was pedagogically intended.

## ACKNOWLEDGEMENTS

The authors thank the University of Maryland Physics Education Research Group, and our collaborators at the University of Colorado (Noah Finkelstein, Katie Hinko, Jessica Hoy, and Doyle Woody).This work is supported by NSF-DUE 1323129.